\documentclass{PoS}

\usepackage{graphicx}
\usepackage{amsmath}
\usepackage{subfigure}

\newcommand*{\chpt}{\raise0.4ex\hbox{$\chi$}PT}
\newcommand*{\schpt}{S\raise0.4ex\hbox{$\chi$}PT}

\newcommand*{\et}{\textit{et al.}}

\newcommand*{\prd}[1]{Phys.\ Rev.\ {D {\bf #1}}}
\newcommand*{\plb}[1]{Phys.\ Lett.\ \textbf{#1 B}}

\newcommand*{\MeV}{{\rm Me\!V}}
\newcommand*{\GeV}{{\rm Ge\!V}}
\newcommand{\fm}{{\rm fm}}

\newcommand{\cM}{\mathcal{M}}

\newcommand{\denom}{16 \pi^2 f^2}
\newcommand{\deltapv}{\delta'_V}
\newcommand{\deltapa}{\delta'_A}
\newcommand{\mxvsq}{m_{X_V}^2}

\newcommand{\mxisq}{m_{X_I}^2}
\newcommand{\myisq}{m_{Y_I}^2}
\newcommand{\muisq}{m_{U_I}^2}

\newcommand{\ftwo}{f}
\newcommand{\mutwo}{\mu}
\newcommand{\Lpptwo}{L''_{(2)}}
\newcommand{\Lptwo}{L'_{(2)}}

\title{SU(2) chiral fits to light pseudoscalar masses and decay constants}
\ShortTitle{SU(2) chiral fits to light pseudoscalar masses and decay constants}

\author{ The MILC Collaboration:}
\author{A.~Bazavov, W.~Freeman and D.~Toussaint\\
        Department of Physics, University of Arizona, Tucson, AZ 85721, USA}

\author{C.~Bernard, \speaker{X.~Du} and J.~Laiho\\
        Department of Physics, Washington University, St.~Louis, MO 63130, USA \\
        E-mail:\email{xiningdu@physics.wustl.edu}}

\author{C.~DeTar, L.~Levkova and M.B.~Oktay\\
	Physics Department, University of Utah, Salt Lake City, UT 84112, USA}

\author{Steven~Gottlieb\\
	Department of Physics, Indiana University, Bloomington, IN 47405, USA}

\author{U.M.~Heller\\
	American Physical Society, One Research Road, Ridge, NY 11961, USA}

\author{J.E.~Hetrick\\
	Physics Department, University of the Pacific, Stockton, CA 95211, USA}

\author{J.~Osborn\\
        Argonne Leadership Computing Facility, Argonne National Laboratory,
Argonne, IL 60439, USA}

\author{R.~Sugar\\
	Department of Physics, University of California, Santa Barbara, CA 93106, USA}

\author{R.S.~Van~de~Water\\
	Department of Physics, Brookhaven National Laboratory, Upton, NY 11973, USA}

\abstract{We present the results of fits to recent asqtad data in the light pseudoscalar sector using SU(2)
partially-quenched staggered chiral perturbation theory. Superfine\ (a\,$\approx$\,0.06\,fm) and ultrafine\
(a\,$\approx$\,0.045\,fm)
ensembles are used, where light sea quark masses and taste splittings are
small compared to the strange quark mass. Our fits include continuum NNLO chiral logarithms
and analytic terms. We give preliminary results for the pion decay constant,
SU(2) low-energy constants and the chiral condensate in the two-flavor chiral limit.}

\FullConference{The XXVII International Symposium on Lattice Field Theory - LAT2009\\
		 July 26-31 2009\\
		 Peking University, Beijing, China}

\begin{document}

\section{Introduction}
At present, most lattice QCD simulations are performed at unphysical light
dynamical quark masses. Fitting of lattice data to forms calculated in chiral
perturbation theory\ (\chpt)~\cite{GL_SU3,GL_SU2} makes possible a controlled
extrapolation of lattice results to the physical light quark masses and to the chiral limit. 
This approach also allows one to determine the values of low-energy constants\ (LECs) in the theory, which are of phenomenological
significance. Although three-flavor \chpt\ has been used successfully for simulations with
2+1 dynamical quarks, we are still interested in the applications of two-flavor
\chpt\ for the following reasons: 

\begin{enumerate}

\vspace{-.3cm}
\item The up and down dynamical quark masses in simulations are usually much smaller than
the strange quark mass, which is near its physical value, hence SU(2) \chpt\ may
serve as a better approximation and probably converges faster than SU(3) \chpt. 

\vspace{-.3cm}
\item Fits to SU(2) \chpt\ can give us direct information about the LECs in the
two-flavor theory, especially $l_3$ and $l_4$. 

\vspace{-.3cm}
\item By comparing results from these two different fits, we can study the systematic errors resulting from the truncations of each version of \chpt.
\vspace{-.3cm}
\end{enumerate}
Recently, some groups have used SU(2) \chpt\ for chiral fits to data from
three-flavor simulations~\cite{HEAVYs_RBC, HEAVYs_PACS-CS}. Here, we perform such an SU(2) chiral analysis for MILC data from simulations with 2+1 flavors of staggered fermions.  

\section{Rooted SU(2) \schpt}
For staggered quarks, the correct effective field
theory is staggered chiral perturbation theory\
(\schpt)~\cite{LEE_SHARPE, AB_M,RUTH_SHARPE_NLO, CB_STAGGERED, BGS_STAGGERED}, 
in which taste-violating effects at finite lattice spacing 
are incorporated systematically. Physical quantities expressed in \schpt\ become
joint expansions in both the quark mass $m_q$ and $a^2$, where
$a$ is the lattice spacing. 

For each quark flavor, there are four species\ (tastes) in the continuum
limit. To obtain physical results, we use the fourth root procedure to
get a single taste per flavor in the continuum limit. Although it has been shown
that this procedure produces violations of locality at non-zero
lattice spacing non-perturbatively~\cite{BGS_NONLOCAL}, recent work indicates that locality and
universality are restored in the continuum limit. For a recent review of the
fourth-root procedure see Ref.~\cite{MILC_REVIEW} and references therein.

In the two-flavor case, only up and down quarks appear in the chiral
theory. Correspondingly, there are only pions, and no kaons, in SU(2) \schpt. 
The staggered Lagrangian is formulated in the same way as in
Ref.~\cite{AB_M}, except that those parts related to the strange quark are omitted.
Following the procedures used in the three-flavor case, one can calculate the
partially-quenched light pseudoscalar mass and decay constant through NLO. The result is 
\cite{DU_SU2}:
\begin{align}
 \frac{m_{P_5^+}^2}{(m_x + m_y)} = &\mutwo \Big\{ 1 + \frac{1}{\denom}\Big(\sum_j R_j^{[2,1]} (\{\cM_{XY_I}^{[2]}\}) l(m_j^2)  \nonumber \\*
                                   &-2a^2 \deltapv \sum_{j}R_j^{[3,1]}(\{\cM_{XY_V}^{[3]}\})
                                   l(m_j^2) + (V\leftrightarrow A) + a^2(\Lpptwo + \Lptwo)\Big) \nonumber \\*
                                   &+\frac{\mutwo}{\ftwo^2}(4l_3+p_1+ 4 p_2)(m_u+m_d) +
				     \frac{\mutwo}{\ftwo^2}(-p_1- 4 p_2)(m_x
+m_y) \Big\},  \label{formula:mpi} 
\end{align}
\begin{align}
f_{P_5^+} = &\ftwo \Big\{ 1 + \frac{1}{\denom} \Big[ -\frac{1}{32}\sum_{Q,B}
l(m_{Q_B}^2) \nonumber \\*      
	    			   &+ \frac{1}{4} \Big( l(\mxisq) + l(\myisq) +
				    (\muisq-\mxisq)\tilde l(\mxisq) + (\muisq - \myisq) \tilde l(\myisq) \Big) \nonumber \\*
             			   &-\frac{1}{2}\Big( R^{[2,1]}_{X_I}(\{\cM^{[2]}_{XY_I}\}) l(\mxisq) +
				    R^{[2,1]}_{Y_I}(\{\cM^{[2]}_{XY_I}\})l(\myisq) \Big) \nonumber \\*
 	 		           &+\frac{a^2 \deltapv}{2}\Big( R^{[2,1]}_{X_V}(\{\cM^{[2]}_{X_V}\}) \tilde l(\mxvsq) + \sum_j D^{[2,1]}_{j,
X_V}(\{\cM^{[2]}_{X_V}\}) l(m_j^2) \nonumber \\*
                                   &+ (X \leftrightarrow Y) + 2\sum_j R^{[3,1]}_{j}(\{\cM^{[3]}_{XY_V}\})l(m_j^2)\Big) + (V \leftrightarrow A)\nonumber \\*
			           &+ a^2(\Lpptwo - \Lptwo) \Big]
+\frac{\mutwo}{2 \ftwo^2}(4l_4 - p_1)(m_u+m_d) + \frac{\mutwo}{2
\ftwo^2}(p_1)(m_x+m_y) \Big\}. \label{formula:fpi}
\end{align}
where $l_3$ and $l_4$ are the standard SU(2) \chpt\ LECs, and $p_1$ and $p_2$ are
two extra NLO LECs that enter in the partially-quenched case. $\deltapv$, $\deltapa$ 
are taste-violating hairpin parameters, and $\Lptwo$, $\Lpptwo$ are taste-violating analytic 
LECs. Chiral logarithms $l(m^2)$, $\tilde l(m^2)$ and residue functions $R$, $D$
are given in Ref.~\cite{AB_M}, with the denominator mass-set arguments in the
SU(2) case defined as:
\begin{align}
\{\cM_{X_V}^{[2]}\} & \equiv \{ m_{X_V}, m_{\eta'_V} \},       &\{\cM_{Y_V}^{[2]}\}  &\equiv \{ m_{Y_V}, m_{\eta'_V} \}, \nonumber \\*
\{\cM_{XY_I}^{[2]}\} & \equiv \{ m_{X_I}, m_{Y_I} \},
&\{\cM_{XY_V}^{[3]}\}  &\equiv \{ m_{X_V}, m_{Y_V}, m_{\eta'_V} \}.
\end{align} 
The numerator mass-set arguments of the residues are always $\{\mu_\Xi\} \equiv
\{m_{U_\Xi}\}$, where the taste label $\Xi$ is taken equal to the taste of the
denominator set.

To Eqs.~(\ref{formula:mpi}) and (\ref{formula:fpi}), we add the NNLO chiral
logarithms that were calculated by Bijnens and L\"ahde~\cite{BIJ_SU2}.
Since taste splittings are not included at NNLO, there is an ambiguity in defining the pion mass in
the continuum formulae. In practice, we use the root mean square \ (RMS) average pion mass in calculations of NNLO chiral logarithms. This is systematic at NNLO as long as the taste splittings between different pions are
significantly less than the pion masses themselves. This condition is best satisfied on the superfine and ultrafine lattices.

\section{Ensembles and Data Sets}
At the present stage, we have the MILC data for the light pseudoscalar mass and decay
constant at five lattice spacings from 0.15\,fm to 0.045\,fm, generated with 2+1
flavors of asqtad improved staggered quarks. For each lattice spacing, we have several different sea quark masses as well as
many different combinations of valence quark masses. In order for the SU(2)
formulae to apply, we require both sea and valence quark masses to be 
significantly smaller than the strange quark mass, {\it i.e.}, $m_\pi^{sea}
\!\ll\! m_K$, and $\;m_\pi^{valence}\! \ll\! m_K$. In the fits described below,
we use the following cutoff on our data sets:
\begin{equation}
m_l \le 0.2 m_s^{phys},\qquad m_x + m_y \le 0.5 m_s^{phys},
\end{equation}
where $m_l$ is the light sea quark mass, and $m_x$ and $m_y$ are the valence masses in
the pion.

To be able to consider the strange quark as ``heavy'' and eliminate
it from the chiral theory, it is also necessary that taste splittings between different
pion states be much smaller than the kaon mass.  Furthermore, taste
splittings should be significantly smaller than the pion mass itself for the continuum
formulae for the NNLO chiral logarithms to be approximately applicable.

The lattices that are at least close to satisfying all these conditions include
four fine (a\,$\approx$\,0.09\,fm) ensembles,
three superfine (a\,$\approx$\,0.06\,fm) ensembles and one 
ultrafine ensemble (a\,$\approx$\,0.045\,fm).  
Relevant parameters for these ensembles
are listed in Table~\ref{table:ensembles}.

\begin{table}
\begin{centering}
\begin{tabular}{|c|c|c|c|c|c|}
\hline
Ensemble	&	$am_l$      &       $am_s$      &       $\beta$    &       size    & $m_\pi L$\\
\hline
\hline
$\approx$ 0.09\,fm (F)	&	0.0062  &       0.031   &       7.09    &       $28^3 \times 96$   &  4.14 \\
$\approx$ 0.09\,fm (F)	&	0.00465 &       0.031   &       7.085   &       $32^3 \times 96$   &  4.10\\
$\approx$ 0.09\,fm (F)	&	0.0031  &       0.031   &       7.08    &       $40^3 \times 96$   &  4.22 \\
$\approx$ 0.09\,fm (F)	&	0.00155 &       0.031   &       7.075   &       $64^3 \times 96$   &  4.80 \\
\hline
\hline
$\approx$ 0.06\,fm (SF)	&	0.0036  &       0.018   &       7.47    &
$48^3 \times 144$   &  4.50 \\
$\approx$ 0.06\,fm (SF)	&	0.0025  &       0.018   &       7.465   &
$56^3 \times 144$   &  4.38 \\
$\approx$ 0.06\,fm (SF)	&	0.0018  &       0.018   &       7.46    &
$64^3 \times 144$   &  4.27 \\
\hline
\hline
$\approx$ 0.045\,fm (UF)	&	0.0028  &       0.014   &       7.81
&       $64^3 \times 192$   &  4.56 \\
\hline
\end{tabular}
\caption{Ensembles used in this analysis. The quantities $am_l$ and $am_s$
are the light and strange sea quark masses 
in lattice units; $m_\pi L$ is the (sea) Goldstone pion mass times 
the linear spatial size. The fine ensembles are not used in our central value
fit, but only in estimating systematic errors.}
\label{table:ensembles}
\end{centering}
\end{table}

 \begin{table}
 \centering
 \begin{tabular}{|c|c|c|c|c|c|}
 \hline
 $a$			 & \multicolumn{2}{|c|}{$\approx$ 0.09\,fm (F)}     &
\multicolumn{2}{|c|}{$\approx$ 0.06\,fm (SF)}    & $\approx$ 0.045\,fm (UF)  \\
 \hline
 $am_S$                 & \multicolumn{2}{|c|}{0.031}     &
\multicolumn{2}{|c|}{0.018}    & 0.014     \\
 \hline
 $am_l$                 & 0.00155          &  0.0062       & 0.0018        &
0.0036          & 0.0028\\
 \hline
 $m_K$(MeV)               & 574          & 613         & 525           & 548          & 565    \\
 \hline
 $m^{\rm Goldstone}_\pi$(MeV) & 177          & 355         & 224           & 317          & 324    \\
 \hline
 $m_\pi^{\rm RMS}$(MeV)	          & 281		 & 416	       & 258	       & 341	      &	334    \\
 \hline
 $m^{\rm I}_\pi$(MeV)     & 346          & 463         & 280           & 359          & 341    \\
 \hline
 \end{tabular}
 \caption{Kaon masses and lightest (sea) pion masses on some sample ensembles.
Here three different pion masses are shown: Goldstone, RMS and singlet.
$r_1=0.3117\,\fm$ is used.}
 \label{table:pionkaonmasses}
 \end{table}

In Table~\ref{table:pionkaonmasses}, we list the Goldstone, RMS and singlet pion
masses on representative ensembles. It can be seen that for the fine
(a\,$\approx$\,0.09\,fm) ensembles, 
either some pion masses are close to the kaon mass, as on ensemble 
$(am_l, am_s) = (0.0062,0.031)$, 
or the taste splittings between pions are comparable to the pion mass, as on ensemble
$(am_l, am_s) = (0.00155,0.031)$. As a result, the data from 
fine lattices may not be well described by SU(2) formulae with
continuum NNLO chiral logarithms. Our central fit uses superfine and
ultrafine data only, while we include fits to all three kinds of lattices to
estimate systematic errors.

There are a total of 29 parameters in our fits. The following list shows how these parameters
are treated in the central fit. 
\vspace{-.1cm}
\begin{itemize}
\vspace{-.1cm}
\item[]{(a)} LO: 2 unconstrained parameters, $\mutwo$ and $\ftwo$. 
\vspace{-.1cm}
\item[]{(b)} NLO (physical): 4 parameters, $l_3$, $l_4$ and two extra LECs $p_1, p_2$ that only appear
in partially-quenched \chpt. All of these parameters are unconstrained.
\vspace{-.1cm}
\item[]{(c)} NLO (taste-violating): 4 parameters. $\deltapv, \deltapa$ are constrained 
within errors at the values determined from SU(3) \schpt\ fits~\cite{MILC04,URS_LAT09}; 
$\Lpptwo$ and $\Lptwo$ are constrained around 0, with width of $0.3$ as 
estimated in Ref.~\cite{MILC04}.
\vspace{-.1cm}
\item[]{(d)} NNLO (physical, ${\cal O}(p^4)$): 5 parameters ($l_1$, $l_2$, $l_7$, $p_3$, $p_4$) that first appear in
meson masses and decay constants in the NNLO chiral logarithms. 
$l_1$ and $l_2$ are constrained by 
the range determined from continuum phenomenology \cite{SU2LEC_VALUES}; $l_7$ is 
not constrained since it is not directly known from 
phenomenology~\cite{SU2LEC_VALUES}. The partially-quenched 
parameters $p_3$ and $p_4$ are not constrained.
\vspace{-.1cm}
\item[]{(e)} NNLO (physical, ${\cal O}(p^6)$): 8 parameters $c_i$,
constrained around 0 with width 1 in ``natural units'' (see Ref.~\cite{MILC04}).
\vspace{-.1cm}
\item[]{(f)} The physical LO and NLO parameters are allowed to vary with lattice spacing by an
amount proportional to $\alpha_s (a\Lambda)^2$, which is the size of the ``generic'' 
discretization errors with asqtad quarks, where $\Lambda$ is some typical
hadronic scale.  This introduces 6 additional parameters that
are constrained around 0 with width corresponding to a scale $\Lambda=0.7\,\GeV$.
\end{itemize}
\vspace{-.1cm}
Alternative versions of the fits, in which the width of the constraints are changed, or some 
constrained parameters are left unconstrained (or {\it vice versa}), have also been tried, 
and the results from those fits are included in the systematic error estimates.

\section{Preliminary Results}
For the central fit, we use three superfine ensembles $(am_l, am_s)
= \{(0.0018, 0.018)$, $(0.0025,\\
0.018)$, $(0.0036,0.018)$\} and one
ultrafine ensemble $(am_l,am_s)\ =\ (0.0028, 0.014)$. 
This fit has a $\chi^2$ of 37 with 33 degrees of freedom, giving a confidence level
${\rm CL}= 0.3$. The volume dependence at NLO has been included in the fit formulae.
A very small ($\le 0.3\%$) correction for ``residual'' finite
volume effects \cite{Bernard:2007ps,Bazavov:2009bb} is applied at
the end of the calculation and incorporated in the systematic errors of our final results.

\begin{figure}
\centering
\subfigure[]{\label{fig:fpi} \includegraphics[width=0.49\textwidth,
angle=0]{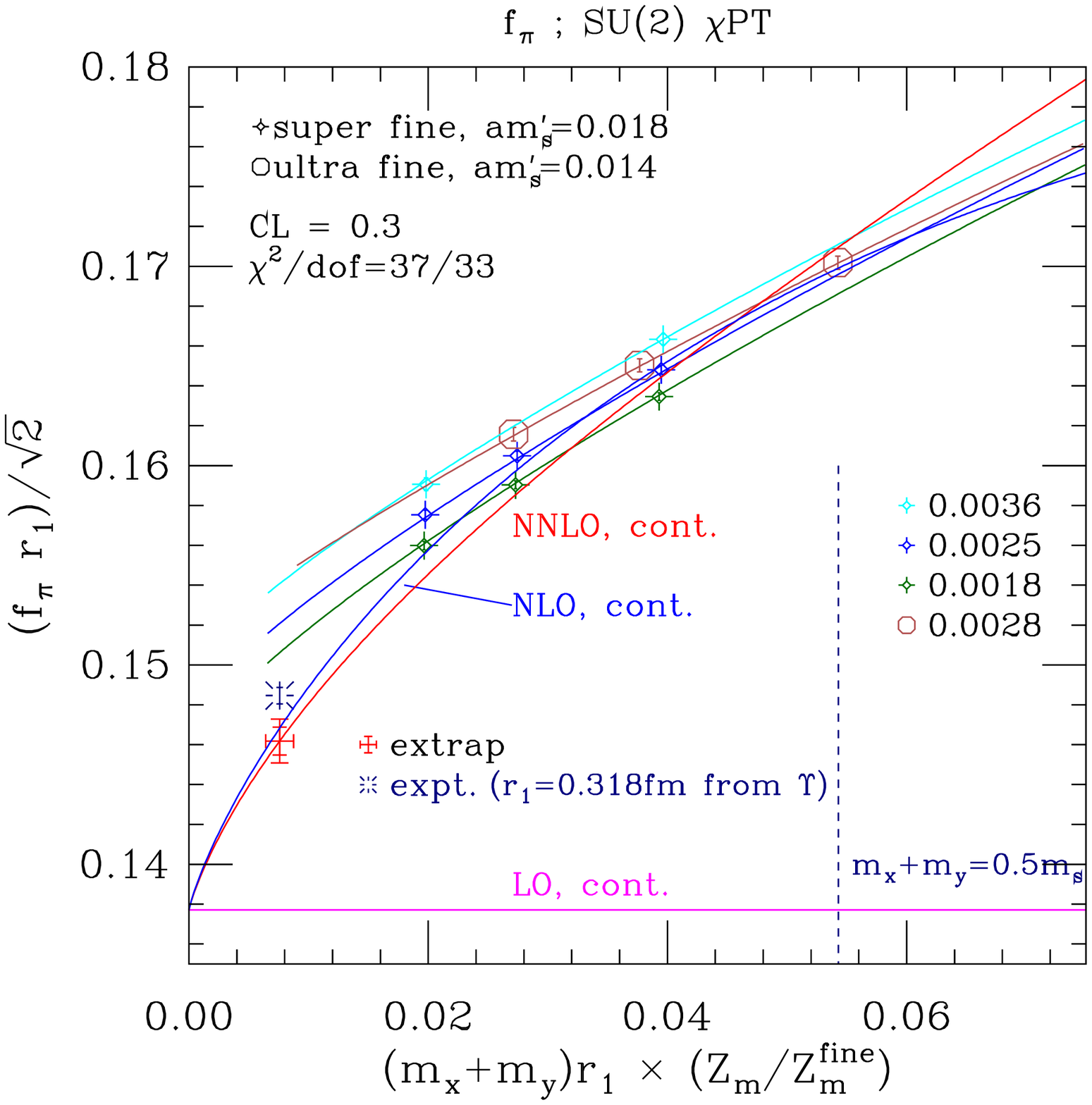}}
\subfigure[]{\label{fig:mpi} \includegraphics[width=0.49\textwidth,
angle=0]{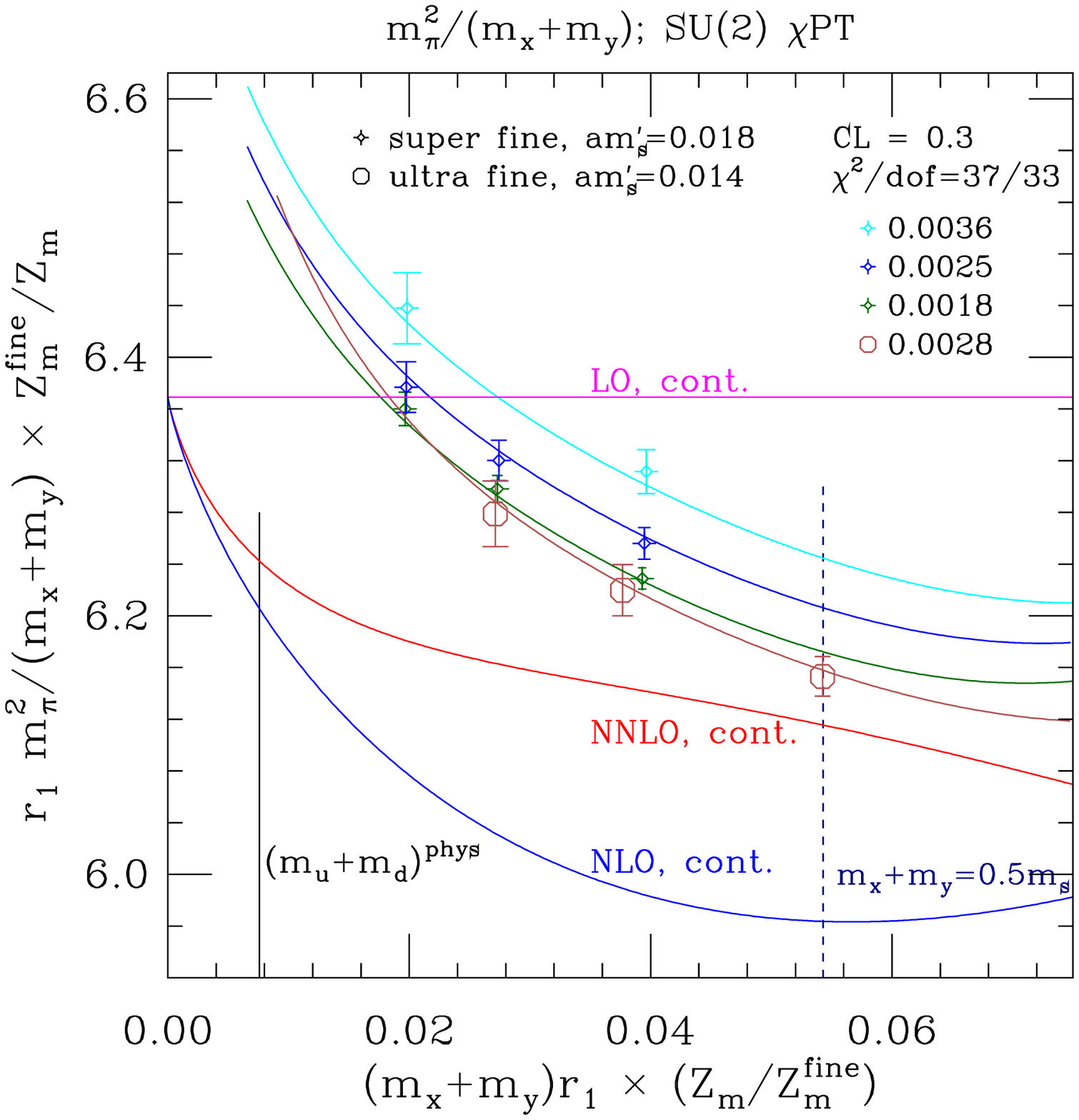}}

\caption{SU(2) chiral fits to $f_\pi$(left) and $m_\pi^2/(m_x+m_y)$(right). Only
points with the valence quark masses equal ($m_x = m_y$) are shown on the plots}

\label{fig:final}
\end{figure}

In Fig.~\ref{fig:final}, we show the fit results for $f_{\pi}$ and $m^2_{\pi}/(m_x + m_y)$
as functions of the sum of the quark masses ($m_x+m_y$). 
The red curves show the complete results through NNLO for full QCD in
the continuum, where we have set taste splitting and taste-violating parameters
to zero, extrapolated physical parameters as $a\to 0$ linearly in
$\alpha_s a^2$, and set valence quark masses and light sea quark masses equal.
Continuum results through NLO and at tree level are shown by blue and magenta
curves, respectively. It can be seen that the convergence of
SU(2) \chpt\ is much better for the decay constant than for the mass. 
Nevertheless, the chiral corrections in both cases appear to be under control.

The SU(2) plots presented previously in Ref.~\cite{URS_CHIRALDYNAMICS09} 
are somewhat different from
those shown here because the earlier fits allowed for $a^2$ variations in the NNLO analytic
parameters ($c_i$). Such variations are of higher order than the 
NNLO terms included in this work. 

At the last step, we find the physical values of the average $u,d$ quark mass 
$\hat{m}$ by requiring
that the $\pi$ has its physical mass, and then find the decay constant
corresponding to this point in Fig.~\ref{fig:final} (left). With the scale parameter
$r_1=0.318(7)\,\fm$~\cite{MILC04}
determined from $\Upsilon$-splittings, we obtain the result for $f_\pi$:
\begin{equation}
f_\pi = 128.3(9)\left(^{+20}_{-8}\right)\,\MeV
\end{equation}
where the first error is statistical and the second is systematic. This agrees
with the PDG 2008 value, $f_\pi = 130.4\pm0.2\,\MeV$~\cite{PDG}.
Alternatively,  using the pion decay constant from NNLO SU(3) \chpt\ fits to define our scale gives
$r_1=0.3117(6)\left(^{+12}_{-31}\right)\,\fm$~\cite{URS_LAT09}. With this new $r_1$, we 
obtain:
\begin{eqnarray}
f_2   = 123.7(9)(18)\,\MeV &\qquad&
B_2   = 2.89(2)\left(^{+3}_{-8}\right)(14)\,\MeV  \nonumber\\   
\bar{l}_3  = 3.0(6)\left(^{+9}_{-6}\right)  &\qquad&
\bar{l}_4  = 3.9(2)(3)    \\
\hat m =  3.21(3)(5)(16)\,\MeV &\qquad&
\langle\bar u u\rangle_2  = -[280(2)\left(^{+4}_{-7}\right)(4)\,\MeV]^3 \nonumber
\end{eqnarray}
The quark masses and chiral condensate are evaluated in the $\overline{\rm MS}$ scheme at
$2\,\GeV$. We use the two-loop renormalization factor in the conversion
\cite{Mason:2005bj}.
Errors from perturbative calculations are listed as the third error in these quantities.
All the quantities agree with results from SU(3) \schpt\ fits~\cite{URS_LAT09}
within errors.

\section{Discussion and Outlook}
We have performed NNLO SU(2) chiral fits to recent asqtad data 
in the light pseudoscalar sector. Results for SU(2) LECs, the pion decay constant, and
the chiral condensate in the two-flavor chiral limit are in good agreement with
those obtained from NNLO SU(3) fits 
(supplemented by higher-order analytic terms for quantities involving strange
valence quarks)\cite{URS_LAT09}.
It can be seen from our plots that SU(2) \chpt\ within its applicable
region converges much faster than SU(3) \chpt. For the point 0.05 on the $x$-axis 
in Fig.~\ref{fig:final}, the ratio of the NNLO correction to the result through NLO
is 0.3\% for $f_\pi$ and 2.6\% for $m_\pi/(m_x + m_y)$. In contrast, the
corresponding numbers in the
SU(3) fits are 2.9\% and 15.6\% respectively (Fig.~2 of Ref.~\cite{URS_LAT09}), although the
large correction in the mass case is partly the result of an anomalously small NLO term.
Note that the SU(3) plots use a non-physical strange quark
mass, $m_s = 0.6 m_s^{phys}$, while for the SU(2) plots, the strange quark mass
is near the physical value, $m_s \approx m_s^{phys}$. This explains why
the two-flavor chiral limits on the SU(3) and SU(2) plots are not the same.

Since the simulated strange quark masses vary slightly between different ensembles, 
the parameters in
SU(2) \schpt\ should also change with ensemble~\cite{DU_SU2}. 
We plan to incorporate this effect in our fit to see if we
can improve the confidence levels. Another step would be to include the
kaon as a heavy particle in SU(2) \schpt\ \cite{Roessl:1999iu} in order to study the physics involving the strange quark,
{\it e.g.}, the kaon mass and decay constant.  This approach has recently been used in Refs.~\cite{HEAVYs_RBC,HEAVYs_PACS-CS}.

\section*{Acknowledgments}
We thank J.\ Bijnens for providing the FORTRAN code to calculate the NNLO
partially-quenched chiral logarithms.


\begin{thebibliography}{99}

\bibitem{GL_SU3}
  J.\ Gasser and H.\ Leutwyler, 
  Nucl.\ Phys.\ B {\bf 250} (1985) 465.

\bibitem{GL_SU2}
  J.\ Gasser and H.\ Leutwyler, 
  Ann. Phys. {\bf 158} (1984) 142.

\bibitem{HEAVYs_RBC}
  C.\ Allton \et\ [RBC-UKQCD Collaboration], Phys.\ Rev.\ D {\bf 78} (2008) 114509 
[arXiv:0804.0473].

\bibitem{HEAVYs_PACS-CS}
  D.\ Kadoh \et\ [PACS-CS Collaboration], arXiv:0810.0351.

\bibitem{LEE_SHARPE}
  W.\ Lee and S.\ Sharpe,
  \prd{60} (1999) 114503.

\bibitem{AB_M}
  C.\ Aubin and C.\ Bernard, 
  \prd{68} (2003) 034014 [hep-lat/0304014] and 074011 (2003) [hep-lat/0306026].

\bibitem{RUTH_SHARPE_NLO}
  S.\ Sharpe, R.\ Van\ de\ Water,
  \prd{71} (2005) 114505 [hep-lat/0409018].

\bibitem{CB_STAGGERED}
  C.\ Bernard,
  \prd{73} (2006) 114503 [hep-lat/0603011].

\bibitem{BGS_STAGGERED}
  C.\ Bernard, M.\ Golterman and Y.\ Shamir,
  \prd{77} (2008) 074505 [arXiv:0712.2560].


\bibitem{BGS_NONLOCAL}
  C.\ Bernard, M.\ Golterman and Y.\ Shamir,
  \prd{73} (2006) 114511 [hep-lat/0604017].

\bibitem{MILC_REVIEW}
  A.\ Bazavov, \et, arXiv:0903.3598, to appear in Rev.\ Mod.\ Phys.

\bibitem{DU_SU2}
  X.\ Du,
  in preparation

\bibitem{BIJ_SU2}
  J.\ Bijnens and T.A.\ L\"ahde,
  \prd{72}, (2005) 074502 [hep-lat/0506004].

\bibitem{MILC04}
  C.\ Aubin \et\ [MILC Collaboration],
  \prd{70}, (2004) 114501 [hep-lat/0407028]. 

\bibitem{URS_LAT09}
  A.\ Bazavov \et\ [MILC Collaboration], these proceedings, PoS {\bf LAT2009} (2009) 079.

\bibitem{SU2LEC_VALUES}
  J.\ Bijnens,
  Prog.\ Part.\ Nucl.\ Phys.{58} (2007) 521 [arXiv:hep-ph/0604043v2]. 

\bibitem{Bernard:2007ps}
  C.~Bernard {\it et al.},
  PoS {\bf LAT2007} (2007) 090 [arXiv:0710.1118 [hep-lat]].

\bibitem{Bazavov:2009bb}
  A.~Bazavov {\it et al.},
  arXiv:0903.3598 [hep-lat].


\bibitem{URS_CHIRALDYNAMICS09}
 A.\ Bazavov \et\ [MILC Collaboration], PoS {\bf CD09} (2009) 007, to appear.

\bibitem{PDG}
  C.\ Amsler \et, 
  \plb{667}(PDG) (2008) 1.

\bibitem{Mason:2005bj}
  Q.~Mason, H.~D.~Trottier, R.~Horgan, C.~T.~H.~Davies and G.~P.~Lepage 
  [HPQCD Collaboration],
  \prd{73} (2006) 114501 [arXiv:hep-ph/0511160].


\bibitem{Roessl:1999iu}
  A.~Roessl,
  Nucl.\ Phys.\ B {\bf 555} (1999) 507 [arXiv:hep-ph/9904230].

\end{thebibliography}
\end{document}